\documentclass{ws-ijmpa}
\begin{document}
\def\prl{{\em Phys. Rev. Lett. }}
\def\jhep{{\em J. High Energy Phys. }}
\def\prc{{\em Phys. Rev. {\bf C} }}
\def\prd{{\em Phys. Rev. {\bf D} }}
\def\jap{{\em J. Appl. Phys. }}
\def\app{{\em Acta Phys. Pol. {\bf B}}}
\def\jpg{{\em J. Phys. G: Nucl.Part. Phys.}}
\def\epjst{{\em Eur. Phys. J.: ST}}
\def\ajp{{\em Am. J. Phys. }}
\def\nima{{\em Nucl. Instr. and Meth. Phys. {\bf A} }}
\def\npa{{\em Nucl. Phys. {\bf A}}}
\def\npb{{\em Nucl. Phys. {\bf B}}}
\def\epjc{{\em Eur. Phys. J. {\bf C}}}
\def\plb{{\em Phys. Lett. {\bf B}}}
\def\mpla{{\em Mod. Phys. Lett. {\bf A}}}
\def\pr{{\em Phys. Rep.}}
\def\zpc{{\em Z. Phys. {\bf C}}}
\def\zpa{{\em Z. Phys. {\bf A}}}
\def\ppnp{{\em Prog. Part. Nucl. Phys.}}

\markboth{R. Sahoo,T.K. Nayak, J. Alam, B.K. Nandi  and S. Kabana}
{Probing the QGP Phase Boundary with Thermal Properties of $\phi$ Mesons} 

%
\catchline{}{}{}{}{}
%

\title{Probing the QGP Phase Boundary with Thermal Properties of $\phi$ Mesons}

\author{Raghunath Sahoo$^{1,2}$\footnote{Corresponding author, e-mail: raghu@mail.cern.ch}, 
Tapan K. Nayak$^{3}$, Jan-e Alam$^{3}$, Basanta K. Nandi$^{4}$ and Sonja Kabana$^{5}$}

\address{ $^{1}$Indian Institute of Technology Indore, Indore-452017, India \\
          $^{2}$Dipartimento di Fisica dell'Universit$\grave{a}$ and Sezione INFN di Padova, Italy \\
          $^{3}$Variable Energy Cyclotron Centre, Kolkata-700064, India \\
          $^{4}$Indian Institute of Technology, Powai, Mumbai-400076, India \\
          $^{5}$SUBATECH, 4, Rue Alfred Kastler, BP 20722 - 44307 Nantes, France}

\maketitle


\begin{abstract}
  A novel attempt has been made to probe the QCD phase boundary  by using the experimental 
data for transverse momenta of $\phi$ mesons produced in nuclear collisions 
at AGS, SPS and RHIC energies. The data are confronted with simple thermodynamic 
expectations and lattice QCD results. The experimental data indicate a first 
order phase transition, with a mixed phase stretching the energy density between 
$\sim$ 1 and 3.2 $GeV/fm^3$ corresponding to SPS energies.
\end{abstract}

\keywords{Heavy-ion collisions, quark-gluon plasma, deconfinement transition} 

\ccode{PACS numbers: 25.75.Nq, 12.38.Mh}

\section{Introduction}
Numerical simulations of Quantum Chromodynamics (QCD) 
on lattice at finite temperature and density predict
that at high temperature and/or energy density hadrons melt down
to a new state of matter, called Quark Gluon Plasma
(QGP) which might have existed after a few microseconds of the 
Big Bang. Lattice QCD based calculations 
\cite{karsch,miller,fodor} indicate that the phase transition 
from a hadronic state to a state of deconfined quarks 
and gluons occurs beyond a critical temperature ($T_c$) 
and a critical baryonic chemical potential  ($\mu^B_c$).
The order of the 
QCD phase transition and the properties of the QGP state 
have been a matter of intense debate, lately.
According to the lattice QCD (lQCD) calculations,
the order of the phase transition depends on the quark masses and 
more importantly, on the baryochemical potential ($\mu_B$). 
These calculations suggest that the QCD phase diagram has
a point in temperature-baryonic chemical potential 
($T-\mu_B$) plane where the first order phase transition ends.
The exact location of this 
point in terms of $T$ and $\mu_B$ is not 
known yet, but the  lQCD calculations suggest that 
it might be within the reach of heavy-ion experiments. 
Since the magnitude of the baryochemical potential 
at the central rapidity region depends on the 
collision energy (it reduces with
the increase in beam energy), by scanning over a broad energy range,
it would be possible to study the change in the 
nature of QCD phase transition along the phase boundary and probe 
the critical point.

A combination of several observables is required in order to 
study the quark-hadron phase boundary and to locate the QCD
critical point. In this article, we point out that $\phi$ meson may be 
used as a  sensitive probe for studying the phase boundary. The $\phi$
meson is the lightest bound state of hidden strangeness ($s\bar{s}$), 
and is produced early in the reaction. The $\phi$ meson is least
suffered by hadronic re-scattering. 
Relatively long lifetime ($\sim 45$ fm/c) of 
$\phi$ mesons may ensure that the decay products are least 
affected by re-scattering with other hadrons as the decay 
occurs outside the fireball~\cite{ko,haglin}. 
On one hand, the theoretical  
calculations~\cite{koch,rafelski} show  different 
amount of modification  of  the $\phi$ meson
properties in the medium due to its interaction with other mesons
in the thermal bath. On the other hand,  the 
results extracted from the analysis of the 
experimental data on the ratio of various hadronic species~\cite{BRS}
indicate that the (inelastic) interactions of $\phi$ with
other hadrons is not significant for temperatures 
below $T_c$. In view of the other systematic
effects entering to the measurements of the 
transverse momentum distribution of $\phi$,  
the change in the properties of $\phi$ in the 
hadronic phase~\cite{koch,rafelski} may not 
be significant enough to alter the results of the present analysis. 
The kinematic observables and flow phenomena of the  
$\phi$ mesons are particularly interesting 
for asserting the equation of state and the nature of 
the phase transition. 

\section{Thermal Properties, Bjorken Energy density and Deconfinement Transition}
\label{sec:2}
The initial energy density plays a very crucial role in the
evolution of the fireball which is estimated as follows. 
The volume of a cylinder of radius $R$ and length $dz$ is
given by $dV=\pi R^2dz=\pi R^2 \tau ~coshy ~d\eta$ where $\tau=\sqrt{t^2-z^2}$,
is the proper time 
and $\eta=\frac{1}{2}~{\mathrm ln}\frac{t+z}{t-z}$,
is the space-time rapidity. Identifying 
$\eta$ with the kinematic rapidity, $y$ and considering the volume element 
of width $dy$ in rapidity space around mid-rapidity
($y=0$) at the initial time, $\tau_0$ one gets the value of the 
initial volume as $dV=\pi R^2\tau_0$. If the total initial energy  
within the rapidity window $dy$ is $dE_T$ then the initial energy density
$\epsilon_{\rm Bj}$, as was first discussed in a seminal paper by
Bjorken \cite{bjorken} is given by,
\begin{equation}
\epsilon_{\rm Bj} 
= \left<\frac{dE_{\rm T}}{dy}\right>\frac{1}{\pi R^2 \tau_0} 
= \left<\frac{dN}{dy}\right>\left<m_{\rm T}\right> \frac{1}{\pi R^2 \tau_0},
\label{BjEqn}
\end{equation}
where $dN/dy$ is the  number of hadrons per unit rapidity and 
$\left<m_{\rm T}\right>$ is the mean transverse mass. 
In the absence of detailed hydrodynamic calculations, this expression
has been widely used by various experiments \cite{starEt,raghuThesis,phenixEt,phenixEt1} to estimate the
initial energy density.  However, the Bjorken energy density given by
Eqn. 1 could be much smaller than the actual initial energy density
produced in nuclear collisions. The estimated energy density in the framework of Bjorken model may be
taken as a lower bound \cite{mg,mga} due to strong reduction, when compared with
the initially produced energy density from longitudinal hydrodynamic work
during the expansion.

In a first order phase transition scenario, the pressure 
increases with temperature until the transition 
temperature $T_c$ is reached, then it remains constant during 
the mixed phase, and continues to increase after the end of 
the mixed phase. In a similar fashion, 
the deconfinement transition can be studied by
observing the variation of average transverse 
momentum ($\left<p_{\rm T}\right>$) of hadrons as a function 
of the hadron multiplicity at mid-rapidity ($dN/dy$)~\cite{vanHove}. 
The $\left<p_{\rm T}\right>$ is expected to reflect the 
average temperature $(T_{\rm th})$ 
and a flow component 
which can be related to the initial pressure. Similarly 
the hadronic multiplicity reflects the entropy density 
of the system. While the purely thermal component of 
$\left<p_{\rm T}\right>$ cannot be related to the initial 
temperature which remains immeasurable directly  above 
$T_c$, however the flow component in the inverse 
slope can reflect the plateau of the pressure during mixed phase.
It has been observed that in central heavy ion 
collisions, the $\left<m_{\rm T}\right>(=\sqrt{p_{\rm T}^2+m^2})$ 
of pions, kaons and protons as a function of $dN_{\rm ch}/dy$ 
shows a step-like behavior as predicted in~\cite{vanHove} 
for a wide range of 
collision energies \cite{bm,marek}. Hydrodynamic 
calculations assuming a first order transition could 
reproduce these data \cite{bm,kodama}. 
Here, we have made an attempt
to make the study for $\phi$-mesons which could provide
a better understanding of the QCD phase boundary, because of
the nature of the interaction of $\phi$ with hadronic  
matter.

The transverse momentum ($p_{\rm T})$ spectra of $\phi$-mesons 
measured for AGS \cite{agsphi,agsphi1}, SPS \cite{na49,na49-1} and RHIC \cite{star,phenix} 
energies have been analyzed in order to obtain the
inverse slope parameter, $T_{\rm eff}$. This effective 
temperature has contributions from both the (random) thermal and the 
collective motions in the transverse direction. It is well known that
the inverse slope ($T_{\rm eff}$) of the transverse momentum spectra
of a hadron of mass $m$ can be related to the `true' freeze-out 
temperature ($T_{\rm th}$) and average radial flow velocity ($<v_{\rm r}>$) at 
the decoupling surface as:
\begin{equation}
T_{\rm eff} = T_{\rm th} + \frac{1}{2}m\left<v_{\rm r}\right>^2,
\label{teff}
\end{equation}
The transverse flow extracted from the $p_{\rm T}$ spectra of $\phi$ meson is 
particularly useful because of its small re-scattering cross-section 
with hadrons, 
the $\phi$ meson is expected to obtain most of its collective flow from 
the partonic phase which is produced in the early stages of heavy-ion 
collisions.

\begin{figure}[pb]
\begin{center}
\includegraphics[scale=0.4]{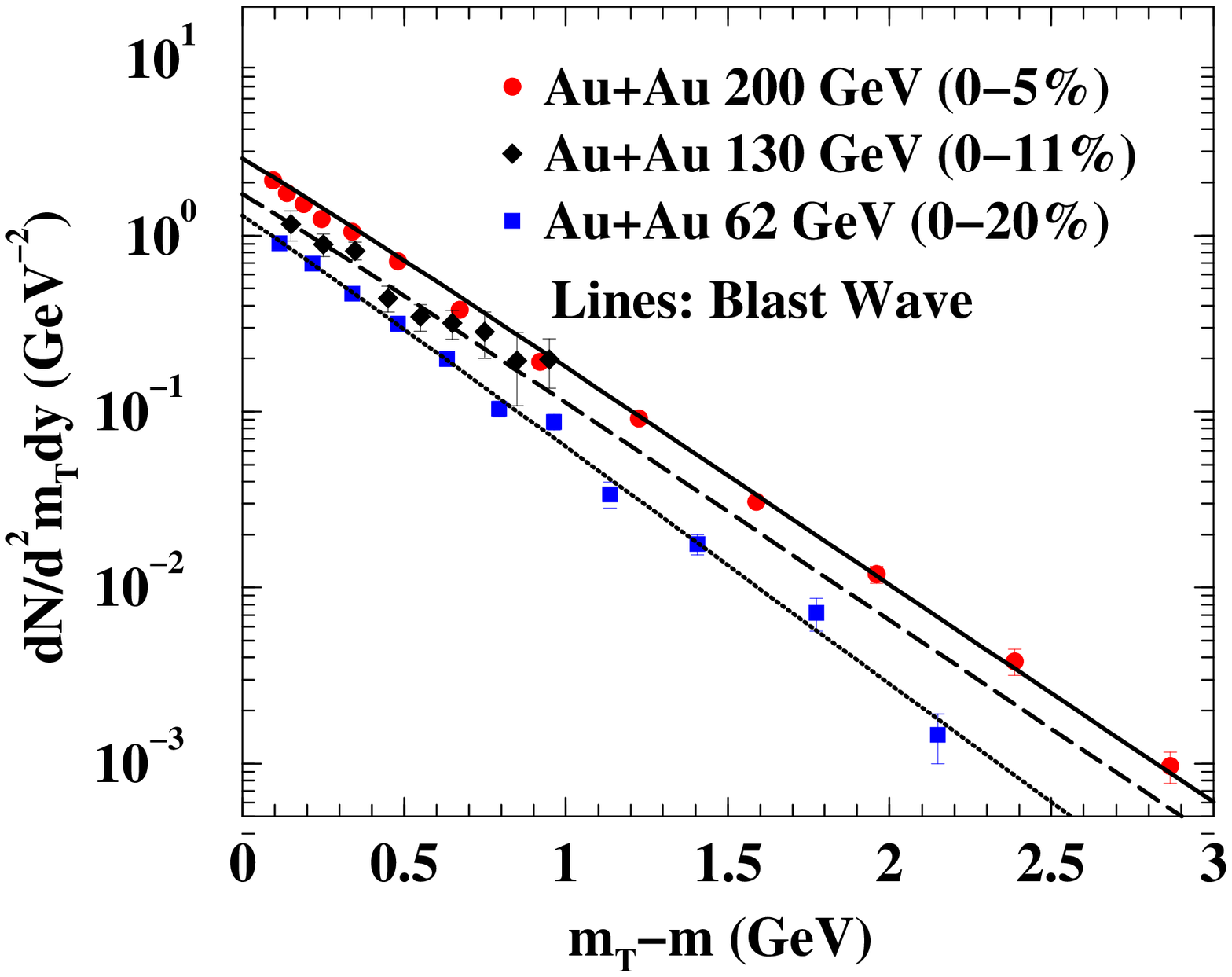}
\includegraphics[scale=0.4]{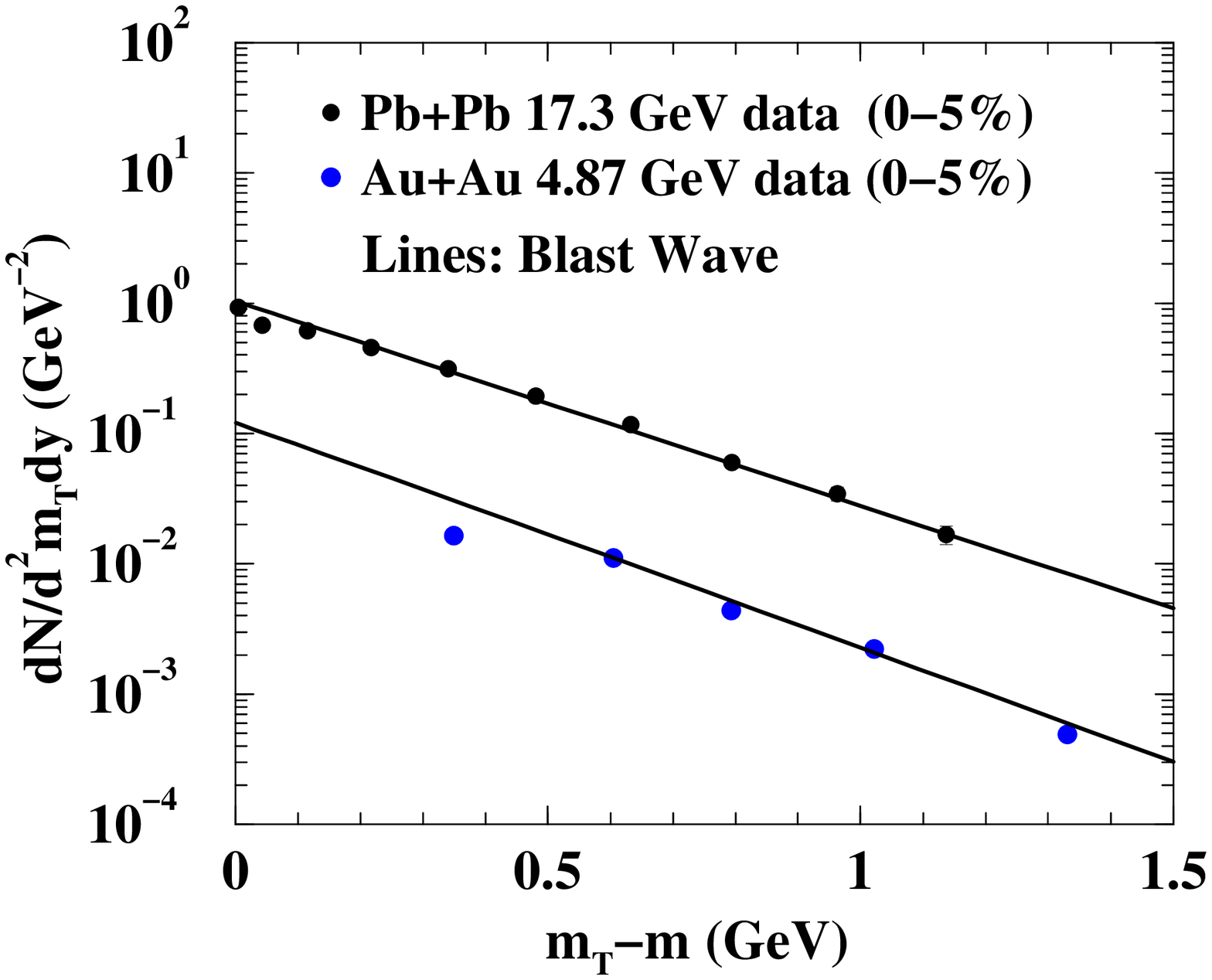}
\caption{
The invariant yield of $\phi$ meson as a function of its
kinetic energy at mid-rapidity for different colliding energies. The top figure
shows the results from the STAR collaboration at RHIC.
The bottom figure shows the results at SPS
and AGS
energies. The theoretical curves are obtained within the 
framework of blast wave model.
}
\label{blastwave}
\end{center}
\end{figure}

Because of long lifetime of $\phi$-meson it can probe the reaction time 
scales in a definite fashion. The $p_{\rm T}$ spectra of $\phi$-mesons 
may reveal quite a few information about the evolution of the system. 
These can be obtained by making a blast wave (BW) model description ~\cite{blastW} 
of the experimental spectra. For this purpose, we have chosen data corresponding to only 
central collisions of \mbox{Au-Au} or \mbox{Pb-Pb} at different
energies. Although the BW parametrization of the $m_{\rm T}$ spectra
could give a set of values for $T_{\mathrm eff}$  and $<v_{\rm r}>$,
we have taken the best possible values (with reduced $\chi^2 \le 1$) for our
present studies.
The results of the BW fit have been shown in Fig.~\ref{blastwave} and the 
extracted values of $<v_{\rm r}>$  and $T_{\mathrm th}$ are listed in Table~I.
The table also enlists $T_{\mathrm eff}$ and other basic parameters of different 
colliding systems at different collision energies.
The similarity of the extracted $T_{\rm th}$ to the critical temperature 
$T_{\rm c}$ predicted by lattice calculations~\cite{karsch} imply that 
the $\phi$ meson freezes out near the phase boundary and could be used 
to extract the properties of QCD matter near the transition point.  
Moreover the large values of $v_{\rm r}$ at RHIC energies indicate that  QGP has undergone 
substantial radial flow.

\begin{table}[h]
\tbl{The table below gives the basic parameters characterizing the colliding systems at different energies, {\it viz.},
centre-of-mass energy, colliding system, centrality, pseudo-rapidity density of charged particles at the central rapidity,
estimated Bjorken energy density, $T_{\rm eff}$, radial flow velocity, thermal temperature. 
The values of $v_{\rm r}$ are obtained from blast wave fit. The estimated error on $v_{\rm r}$ is about 10\%.}
{\begin{tabular}{@{}ccccccccc@{}} \toprule
$\sqrt{s_{\rm NN}}$($GeV$) & System & Centrality & $\frac{dN_{\rm ch}}{d\eta}$ & $\epsilon_{\rm Bj}$($GeV/fm^3$)    
&     $T_{\rm eff}$($GeV$)      &  $v_{\rm r}$   & $T_{\rm th}$($GeV$)            &   Ref.         \\
\hline
 4.87  &  \mbox{Au-Au} & 0-5\%   &  120  &    1.3  & 0.196$\pm$0.02$\pm$0.02  & 0.39   & 0.118   & \cite{agsphi,agsphi1}  \\
 6.3   &  \mbox{Pb-Pb} & 0-7.2\% &  226  &    1.0  & 0.196$\pm$0.02$\pm$0.02  & 0.43   & 0.102    & \cite{na49,na49-1}    \\
 7.6   &  \mbox{Pb-Pb} & 0-7.2\% &  251  &    1.05 & 0.237$\pm$0.18$\pm$0.02  & 0.43   & 0.143   & \cite{na49,na49-1}    \\
 8.7   &  \mbox{Pb-Pb} & 0-7.2\% &  272  &    1.1  & 0.244$\pm$0.09$\pm$0.06  & 0.43   & 0.150    & \cite{na49,na49-1}    \\
12.3   &  \mbox{Pb-Pb} & 0-7.2\% &  340  &    1.25 & 0.240$\pm$0.08$\pm$0.01  & 0.43   & 0.145   & \cite{na49,na49-1}    \\
17.3   &  \mbox{Pb-Pb} & 0-5\%   &  426  &    2.9  &
0.299$\pm$0.07$\pm$0.01  & 0.5    & 0.171     & \cite{na49,na49-1}    \\
62.4   &  \mbox{Au-Au} & 0-20\%  &  472  &    3.5  & 0.328$\pm$0.06$\pm$0.02  & 0.55   & 0.173  & \cite{star}    \\
130    &  \mbox{Au-Au} & 0-11\%  &  520  &    4.7  & 0.379$\pm$0.05 $\pm$0.05 & 0.63   & 0.176  & \cite{star}    \\
200    &  \mbox{Au-Au} & 0-10\%  &  670  &    4.9  & 0.359$\pm$0.05$\pm$0.02  & 0.63   & 0.157  & \cite{star}    \\
200    &  \mbox{Au-Au} & 0-10\%  &  670  &    5.4  & 0.376$\pm$0.024$\pm$0.02 & 0.63   & 0.174  & \cite{phenix}  \\
\botrule
\end{tabular}\label{tab1}}
\end{table}

\begin{figure}
\begin{center}
\includegraphics[width=3.1in]{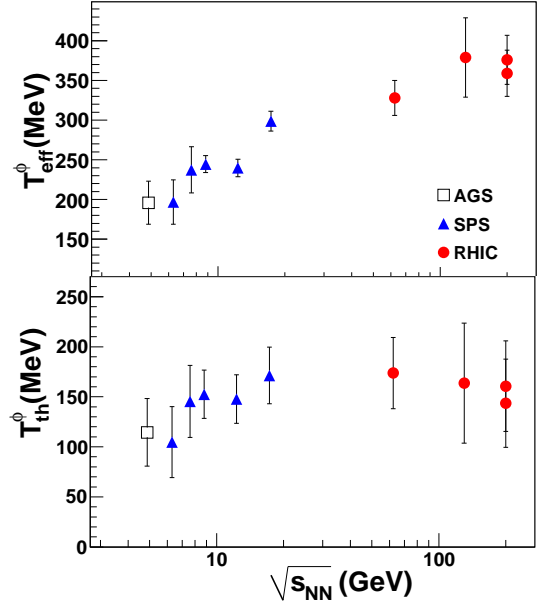}
\caption{
Top panel shows the effective temperature of $\phi$-meson
($T_{\textrm{eff}}^{\phi}$) and the bottom panel shows the
$T_{\textrm{th}}^{\phi}$ as a function of center-of-mass 
energy ($\sqrt{s_{\rm NN}}$) from AGS, SPS to RHIC energies.}
\label{withCoM}
\end{center}
\end{figure}

A compilation of measured data for central heavy ion collisions for
$T_{\rm eff}$, for low $p_{\rm T}$ range ($0 < p_{\rm T} \leq 3.0 $ GeV/c) 
and at mid-rapidity has been 
made~\cite{agsphi,agsphi1,na49,na49-1,star,phenix,floris,na38,na50} as shown in Table~I. 
The results are depicted in Fig.~\ref{withCoM} as a function of collision 
energy $\sqrt{s_{\rm NN}}$. Both the hadronic and leptonic decay channels of 
$\phi$-mesons are included in this figure. 
The $T_{\rm th}$ and $v_r$  in Eq. (\ref{teff}) 
have been obtained from the BW  description of the transverse
momentum spectra of $\phi$. 
It is necessary to mention here
that even at top RHIC energies, it has been observed \cite{starPLB} that 
a change of impact parameter of the collision doesn't make the required 
change of energy density to observe a change in phase of the produced
matter (no step-like behavior in $<p_{\rm T}>$ has been seen). So it is imperative 
to study the variation of temperature with energy density or center of mass 
energy for a fixed collision centrality.

\begin{figure}
\begin{center}
\includegraphics[width=3.1in]{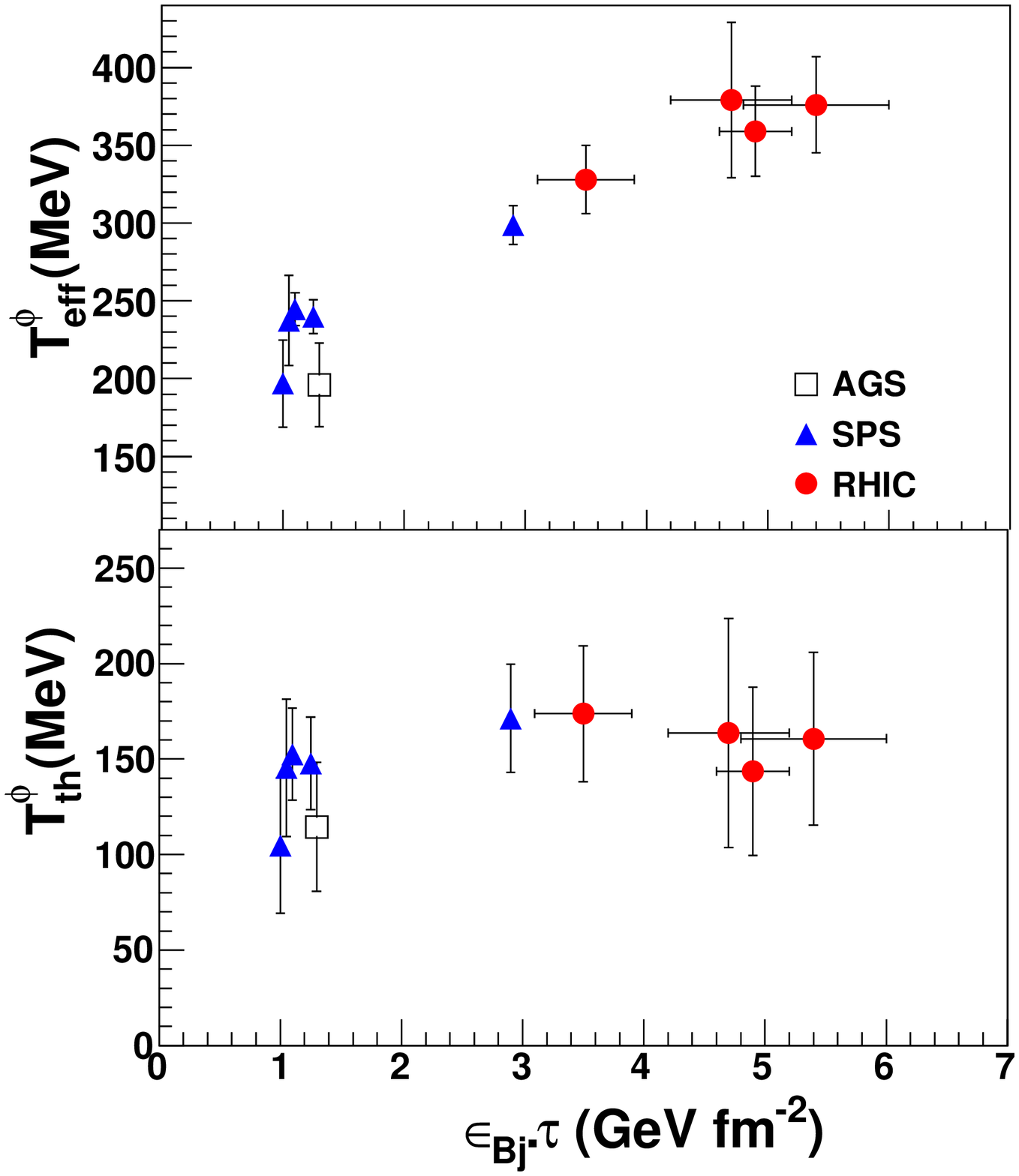}
\caption{ Top panel shows the effective temperature of $\phi$-meson
($T_{\textrm{eff}}^{\phi}$)
and the bottom panel shows the
$T_{\textrm{th}}^{\phi}$
as a function of energy density multiplied to formation time 
($\epsilon_{\rm Bj}.\tau$) from AGS, SPS to RHIC energies.
}
\label{withEDensity}
\end{center}
\end{figure}

From the experimental measurements of $dE_{\rm T}/dy$ or $dN_{\rm ch}/dy$ 
with $\left<m_{\rm T}\right>$, $\epsilon_{\rm Bj}.\tau$ can be estimated for all 
centralities, colliding systems and collision energies. In the absence of 
any accurate estimate of the formation time $\tau$, we use the quantity of 
$\epsilon_{\rm Bj}.\tau$.  
Figure \ref{withEDensity} shows $T_{\textrm{eff}}^{\phi}$ (top) and 
$T_{\textrm{th}}^{\phi}$ (bottom) of $\phi$-mesons as functions of 
$\epsilon_{\rm Bj}.\tau$. In both the figures (\ref{withCoM} and \ref{withEDensity}),
the inverse slope parameter of the $p_T$-spectra, 
$T_{\textrm{eff}}^{\phi}$, shows an increase with collision energy/energy 
density till lower SPS energy is reached. 
A further increase in collision energy leads to 
an increase of early temperature and pressure of the system. As a 
consequence the transverse momenta of the produced hadrons and hence the
inverse slope parameters increase with collision energy. This is followed by
the region of constant temperature which corresponds to higher SPS energies,
where it is expected that the system undergoes a deconfinement transition
with the creation of a co-existing phase of partons and hadrons, 
which is signaled by a plateau-like structure in the above spectrum. 
The resulting modification of Equation of State (EoS)
suppresses the transverse expansion. 
This observation of a plateau-like structure is equivalent to a liquid-gas 
phase transition with involvement of latent heat of the system, characteristic of  
a first order phase transition. Thus the experimental data indicate a first
order phase transition, with a mixed phase stretching energy density between
$\sim 1$ and $3.2 ~ $GeV/fm$^3$. At higher energies (corresponding to RHIC 
energies), $T_{\textrm{eff}}^{\phi}$ further increases with collision energy.
The EoS at the early stage becomes stiff again leading to increase in 
early pressure and temperature. The thermal component, 
$T_{\textrm{th}}^{\phi}$, 
also shows a plateau like behavior in going toward the RHIC 
energies. The extraction of $T_{\textrm{th}}^{\phi}$ involves the flow 
component, $v_{\rm r}$, which has an added significance. As a function of 
collision energy, $v_{\rm r}$ shows similar behavior as 
$T_{\textrm{eff}}^{\phi}$~\cite{nu}. A similar pattern is suggested for the 
elliptic flow as a function of collision energy~\cite{voloshin}. 
The observation of a mixed phase goes in line with the fact that excitation 
function of various observables show 
anomalous behavior or saturation effects starting at lower SPS energies 
\cite{raghu,raghu-1,raghu-2,alt,sv}. This could be related to the onset of deconfinement 
corresponding to SPS energy regime. In addition, the fact that microscopic 
transport models, based on hadronic degrees of freedom failed to reproduce 
the observed behavior of kaon inverse slope \cite{el,el-1,wagner}, also indicated 
the creation of a deconfinement transition corresponding to this energy 
regime. It has also been observed that the inverse slope parameter in $p+p$
interactions increases smoothly with collision energy with the absence of
any plateau-like structure which signify a first order phase transition
\cite{alt}. An interesting review on onset of deconfinement in
nucleus-nucleus collisions has recently been done by Gazdzicki {\it et al.} \cite{onset}.

\section{Summary}
\label{sec:4}
In summary, we observe the anomalous energy (density) dependence of
transverse mass spectra of $\phi$-mesons produced in central heavy ion
collisions. The inverse slope of the $m_{\rm T}$ spectrum shows an increase 
in AGS and RHIC energy regime, whereas it is almost constant in the 
intermediate SPS energies. This observation goes in line with the previous
observations for lighter mesons. This anomaly could be caused by the
modification of the equation of state in the transition region of the
QCD deconfinement transition. The present analysis also indicates 
a first order phase transition stretching over an energy density 
of $1-3.2 ~$ GeV/fm$^3$, which agrees with 
the lQCD prediction for the critical energy density 
($\epsilon_c \sim 0.7 ~$ GeV/fm$^3$) for deconfinement.
This observation is very exciting in view of the observed anomalies in the
excitation functions of other observables, namely strangeness production,
in view of the onset of deconfinement in the SPS energy domain. 
It would have been equally interesting to make a similar study on the
spectra of $\Omega$ hyperons, as it is sensitive to the matter
equation of state at the early stages of collisions. However, at
present, enough data are not available for the $\Omega$ hyperons to
make a similar study. The ongoing 
RHIC beam energy scan and the same at SPS will certainly help in getting 
a clear picture to this problem, with the advent of new data sets in the
intermediate energy regime. \\

\noindent {\bf Acknowledgments : }{We thank Drs. D. Jouan, M. Floris, V. Friese, 
A. De Falco for providing SPS data and Profs. Y.P. Viyogi and D.P. Mahapatra for useful discussions.}


\begin{thebibliography}{0}    

\bibitem{karsch} F. Karsch, \ppnp{\bf 62}, 503 (2009).

\bibitem{miller} D.E. Miller, \pr{\bf 443}, 55 (2007).

\bibitem{fodor} Z. Fodor and S. Katz, \jhep{\bf 04}, 050 (2004).

\bibitem{ko} C.M. Ko and D. Seibert, \prc{\bf 49}, 2198 (1994).

\bibitem{haglin} K. Haglin, \npa{\bf 584}, 719 (1995).

\bibitem{koch} L. Alvarez-Ruso and V. Koch, \prc{\bf 65}, 054901 (2002).

\bibitem{rafelski} I. Kuznetsova and J. Rafelski, \prc{\bf 82}, 035203 (2010).

\bibitem{BRS} P. Braun-Munzinger, K. Redlich and J. Stachel,
Quark Gluon Plasma 3, Eds. R. C. Hwa and X. N. Wang, World Scientific (2003).

\bibitem{bjorken} J.D. Bjorken, \prd{\bf 27}, 140 (1983).

\bibitem{starEt} J. Adams {\it et al.} (STAR Collaboration), \prc{\bf
   70}, 054907 (2004).

\bibitem{raghuThesis} Raghunath Sahoo (STAR Collaboration),
  Ph.D. Thesis (2007), arXiv: 0804.1800 [nucl-ex].

\bibitem{phenixEt} K. Adcox {\it et al.} (PHENIX Collaboration), \prl{\bf 87}, 052301 (2001).

\bibitem{phenixEt1} S.S. Adler {\it et al.} (PHENIX Collaboration), \prc{\bf 71}, 034908 (2005).

\bibitem{mg} M. Gyulassy and T. Matsui, \prd{\bf 29}, 419 (1984).

\bibitem{mga} A. Dumitru and M. Gyulassy, \plb{\bf 494}, 215 (2000).

\bibitem{vanHove} L. Van Hove, \plb{\bf 118}, 138 (1982).

\bibitem{bm} B. Mohanty {\it et al.}, \prc{\bf 68}, 021901 (2003).

\bibitem{marek} M.I. Gorenstein, M. Gazdzicki and K.A. Bugaev, \plb{\bf 567}, 175 (2003).

\bibitem{kodama} Y. Hama {\it et al}, \app{\bf 35}, 179 (2004).

\bibitem{agsphi} Y. Akiba {\it et al.} (E-802 Collaboration), \prl{\bf
    76}, 2021 (1996).

\bibitem{agsphi1}  B.B. Back {\it et al.} (E917 Collaboration), \prc{\bf 69}, 054901 (2004).

\bibitem{na49} C. Alt {\it et al.} (NA49 Collaboration), \prl{\bf 94},  052301 (2005).

\bibitem{na49-1} C. Alt {\it et al.} (NA49 Collaboration), \prl{\bf 78}, 044907 (2008). 

\bibitem{star} B.I. Abelev {\it et al.} (STAR Collaboration), \prc{\bf 79}, 064903 (2009).

\bibitem{phenix} S.S. Adler {\it et al.} (PHENIX Collaboration), \prc{\bf 72}, 014903 (2005).

\bibitem{blastW} E. Schnedermann, J. Sollfrank  and U. Heinz, \prc{\bf 48}, 2462 (1993). 

\bibitem{floris} M. Floris {\it et al.} (NA60 Collaboration), \epjc{\bf 49}, 255 (2007).

\bibitem{na38} M.C. Abreu  {\it et al.} (NA38 Collaboration), \plb{\bf 368}, 239 (1996).

\bibitem{na50} B. Alessandro {\it et al.} (NA50 Collaboration), \plb{\bf 555}, 147 (2003).

\bibitem{starPLB} B.I. Abelev {\it et al.} (STAR Collaboration), \plb{\bf 673}, 183 (2009).

\bibitem{nu} N. Xu, \npa {\bf 751}, 109c (2005).

\bibitem{voloshin} S.A. Voloshin, A.M. Poskanzer and R. Snellings; arXiv:0809.2949 [nucl-ex].

\bibitem{raghu} J. Cleymans {\it et al.}, \plb{\bf 660}, 172 (2008).

\bibitem{raghu-1} J. Cleymans {\it et al.}, \epjst{\bf 155}, 13 (2008).

 \bibitem{raghu-2} J. Cleymans {\it et al.}, \jpg{\bf 35}, 104147 (2008).

\bibitem{alt} C. Alt {\it et al.} (NA49 Collaboration), \prc{\bf 77}, 024903 (2008).

\bibitem{sv} S.V. Akkelin and  Yu M. Sinyukov, \prc{\bf 73}, 034908 (2006).

\bibitem{el} E.L. Bratkovskaya {\it et al.}, \prc{\bf 69}, 054907 (2004).

\bibitem{el-1} E.L. Bratkovskaya {\it et al.}, \prl{\bf 92}, 032302 (2004).

\bibitem{wagner} M. Wagner, A.B. Larionov, and U. Mosel, \prc{\bf 71},
  034901 (2005).

\bibitem{onset} M. Gazdzicki  {\it et al.}, \app {\bf 42}, 307 (2011).






\end{thebibliography}
\end{document}